\RequirePackage{lineno}
\documentclass[twocolumn,showpacs,superscriptaddress,amsmath,amssymb,nofootinbib]{revtex4-2}
\usepackage{graphicx}
\usepackage{subfigure}
\usepackage{subcaption}
\usepackage{epsfig}
\usepackage{overpic}
\usepackage{dcolumn}
\usepackage{ulem}
\usepackage{makecell}
\usepackage{bm}
\usepackage{color}
\usepackage{lineno}
\usepackage{xspace}
\usepackage{multirow}
\usepackage{epstopdf}
\usepackage{xcolor}
\usepackage{soul}
\usepackage{verbatim}
\usepackage{enumitem}
\usepackage{todonotes}
\usepackage{cancel}
\usepackage[toc,page]{appendix}
\usepackage[colorlinks,allcolors=black]{hyperref}
\allowdisplaybreaks[4]
\usepackage{caption}
\usepackage{diagbox}

\newcommand{\iqm}{\affiliation{State Key Laboratory of Nuclear Physics and Technology, Institute of Quantum Matter, South China Normal University, Guangzhou 510006, China}}

\newcommand{\moe}{\affiliation{Key Laboratory of Atomic and Subatomic Structure and Quantum Control (MOE), Guangdong-Hong Kong Joint Laboratory of Quantum Matter, Guangzhou 510006, China}}

\newcommand{\scnt}{\affiliation{Southern Center for Nuclear-Science Theory (SCNT), Institute of Modern Physics, Chinese Academy of Sciences, Huizhou 516000, Guangdong Province, China}}

\newcommand{\gbrc}{\affiliation{Guangdong Basic Research Center of Excellence for Structure and Fundamental Interactions of Matter, Guangdong Provincial Key Laboratory of Nuclear Science, Guangzhou 510006, China}}

\newcommand{\OU}{\affiliation{Research Center for Nuclear Physics (RCNP), Osaka University, Ibaraki 567-0047, Japan}}

\newcommand{\JAEA}{\affiliation{Advanced Science Research Center, Japan Atomic Energy Agency (JAEA), Tokai 319-1195, Japan}}

\newcommand{\Nishogakusha}{\affiliation{Nishogakusha University, 6-16, Sanbancho, Chiyoda, Tokyo 102-8336, Japan}}

\newcommand{\Keio}{\affiliation{Research and Education Center for Natural Sciences, Keio University, Hiyoshi 4-1-1, Yokohama, Kanagawa 223-8521, Japan}}

\newcommand{\SKCMM}{\affiliation{International Institute for Sustainability with Knotted Chiral Meta Matter (SKCM$^2$), Hiroshima University, 1-3-2 Kagamiyama, Higashi-Hiroshima, Hiroshima 739-8511, Japan}}

\begin{document}
\include{def-com}
\title{\boldmath Finite-temperature effects on the threshold cusps in $\pi\pi$ and $D\bar{D}^{\ast}$ scatterings from relativistic heavy-ion collisions}

\author{Ying Zhang}
\email{yingzhang@m.scnu.edu.cn}
\email{ying@rcnp.osaka-u.ac.jp}
\iqm
\OU
\moe
\gbrc
\author{Atsushi Hosaka}
\email{hosaka@rcnp.osaka-u.ac.jp}
\OU
\JAEA
\author{Qian Wang}
\email{qianwang@m.scnu.edu.cn}
\iqm
\OU
\moe
\scnt
\author{Shigehiro Yasui}
\email{yasuis@keio.jp}
\Nishogakusha
\Keio
\SKCMM
\date{\today}

\begin{abstract}
We investigate how the temperature influences the threshold cusps in
meson-meson scatterings, i.e., 
$\pi\pi$ and $D\bar{D}^\ast$ (or $D^{\ast}\bar{D}$) scatterings, using the production rates and propagators obtained at finite temperature. The lineshape of production rate of $\pi\pi$ at different temperatures demonstrates that the cusp structure in $\pi\pi$ scattering is mildly enhanced as the temperature increases. As for the $\pi\pi$ propagator, which includes the isospin symmetry breaking, its lineshape displays a unique plateau-like structure and this structure will also be enhanced as the temperature increases.
For comparison, the lineshape of the $D\bar{D}^\ast$ propagator 
including the isospin symmetry breaking is also investigated at different temperatures. As the temperature increases, its lineshape shows a similar plateau-like structure but with some different properties when the temperature modifications to the masses and widths of $D$ and $\bar{D}^\ast$ are considered.
\end{abstract}
\maketitle

\section{Introduction}\label{sec:Introduction}

The study of threshold cusp has a long history and can be traced back to decades ago \cite{Wigner:1948zz, Breit:1957zz}. It is produced by the square-root branch points at a two-body threshold and it sometimes may bring confusions to the identification of a resonance close to the threshold of two particles because of the low energy resolution. However, it may be useful for working out some physical problems. It had been pointed out very early by Budini and Fonda \cite{Budini:1961bac} that these cusps might be used to investigate $\pi\pi$ scattering and extract a combination of scattering lengths because the strength of the cusp is proportional to the charge-exchange $\pi\pi$ scattering amplitude at the threshold. Until now, to determine the combination of scattering lengths, most of the researches about the threshold cusp in $\pi\pi$ scattering are conducted using the theories in vacuum \cite{NA482:2005wht, Cabibbo:2004gq, Batley:2009ubw, Gamiz:2006km, Kubis:2009vu}, such as the chiral perturbation theory and non-relativistic effective field theory. However, it is known that the study of hadron properties at finite temperature in relativistic heavy ion collisions (HICs) provides further useful information on the strong interaction dynamics. So studying the threshold cusp in $\pi\pi$ scattering at finite temperature may provide us with more information on the $\pi\pi$ scattering and strong interaction. 

In fact, there are already a few researches studying the temperature effect on another kind of kinematical effect, i.e., the triangle singularity, by analyzing the behavior of the one-loop three-point function at different temperatures~\cite{Abreu:2020jsl,Abreu:2021xpz,Llanes-Estrada:2021jud,Llanes-Estrada:2021ath}. These studies also consider the temperature dependence of mass and width of the intermediate particles in their analysis. Based on the computations in thermal field theory, they arrive at the same conclusion that the triangle singularities will be significantly suppressed in the environment with the sufficiently high temperature. Here we turn to focus on the threshold cusp behavior in $\pi\pi$ scattering at different temperatures, which may deepen our understanding of the threshold cusp. For comparison, the temperature effect on the threshold cusp in $D\bar{D}^\ast$ (or $D^\ast\bar{D}$) scattering is also investigated.

The paper is organized as follows. In Sec.~\ref{sec:two_point} we present how the temperature affects the one-loop two-point function. Then in Sec.~\ref{sec:application1} and Sec.~\ref{Production}, we investigate the temperature effect on the threshold cusp by virtue of the amplitude, production rate, and propagator in $\pi\pi$ scattering. In the end of this section, we briefly comment on electromagnetic effects.
In Sec.~\ref{sec:application2}, for comparison, the temperature effect on the threshold cusp in the $D\bar{D}^\ast$ meson system is also studied with the propagator of $D\bar{D}^\ast$ scattering.
Sec.~\ref{sec:sum} is devoted to our summary and outlook.
In Appendix~\ref{sec:Sommerfeld-Watson}, we summarize the Sommerfeld-Watson transformation with Matsubara frequencies at finite temperature.
In Appendix~\ref{sec:2pft}, we present the calculation of one-loop two-point function at finite temperature. In Appendix \ref{Tfull&Tper}, we make a comparison between the scattering amplitude $T$ in full order ( $T^{\text{full}}$) and in perturbation order ( $T^{\text{pert}}$). The interaction effect on the scattering amplitude $T$ is also analyzed.
In Appendix~\ref{sec:isobreaking}, we show the calculation of the propagator with the isospin symmetry breaking. 

Some caveats are in order. The cusp is the phenomenon at the threshold. Therefore, for the pion scattering, 
one would perform systematic calculation of chiral perturbation theory. 
In fact, perturbation calculation has been performed previously~\cite{Kaiser:1999mt,GomezNicola:2002tn},  where they have investigated the scattering length with the inclusion of all possible one-loop diagrams, $s$, $t$, $u$ and tad poles.  It is important to include all of them for systematic studies based on perturbation theory. 
Now for the cusp, it is the singularity in the $s$-channel process (Cutkosky cut) which is primarily driven by the loop in the $s$-channel, that is the $I_2$-function as we will define later.  Furthermore, in general, the shape and strength of the cusp may be modified by the interaction and temperature. Therefore, in the present paper we construct the scattering matrix $T$ by summing up the $s$-channel loop to all orders. Our formula can be applied to general cases of
particles with various masses and with stronger interactions than those applicable to perturbative calculations. This strategy is also reasonable for applications of the present method to such as $D \bar {D}^*$ scattering which we will briefly discuss using the loop function $I_2$ in the present paper.

\section{One-loop two-point function at finite temperature: general case}\label{sec:two_point}

\begin{figure}[h]
    \centering
    \includegraphics[width=1\linewidth]{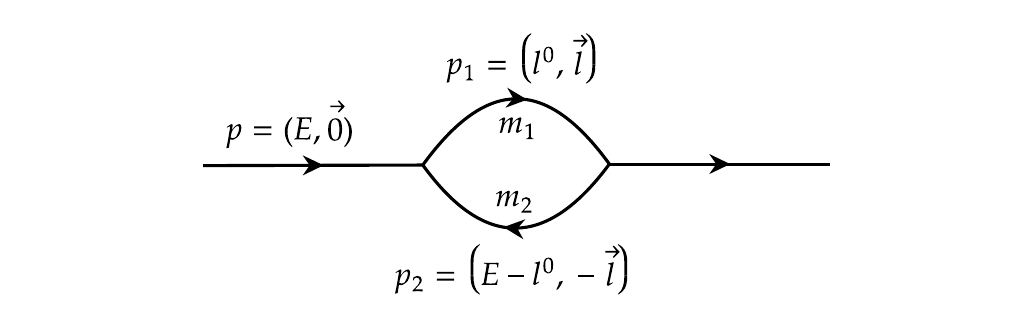}
    \caption{The loop diagram with two internal particles 1 and 2.}
    \label{fig:two_point}
\end{figure}

In this section, we focus on the one-loop two-point function to discuss the two-body threshold cusp at finite temperature in general two-meson systems. We adopt the rest frame of the center-of-mass of the two particles, and set the four-dimensional 
 energy-momentum $p = (E, \vec{0})$ in total as shown in Fig.~\ref{fig:two_point}. 
We consider two bosons of spin zero which are labeled by 1 and 2, respectively.

The scalar one-loop two-point function in relativistic form reads
\begin{align}
&I_2(E,m_1,m_2)\notag\\
=&~i\int\frac{d^4l}{(2\pi)^4}\frac{1}{(l^{02}-w_1^2+i\epsilon)[(E-l^0)^2-w_2^2+i\epsilon]},
\label{eq:two_point_0}
\end{align}
in the rest frame.
In the loop function, $w_i=\sqrt{\vec{l}\,^2+m_i^2}$ ($i=1,2$) is introduced where the energy-momentum is denoted by $l=(l^{0},\vec{l})$.
Here, $E$ is the initial energy, $m_1$ and $m_2$ are masses of particle 1 and 2, respectively, and $\epsilon$ is an infinitely small positive number.

It is known that, at finite temperature, the definition of the partition function for boson system requires the fields to be periodic in imaginary time~\cite{Kapusta:2007xjq, Kapusta:1989ca,Bellac:2011kqa,Nieto:1993pr}. Thus, the zeroth component of the loop momentum $l^0$ is replaced by Matsubara frequencies 
$i\omega_n=i2\pi n/\beta$ ($n\in \textbf{Z}$ and $\beta=1/T$, the inverse temperature $T$) for the intermediate mesons \cite{ Kapusta:2007xjq,Kapusta:1989ca,Bellac:2011kqa, Nieto:1993pr, Aurenche:1991hi, Evans:1991ky, Laine:2016hma}.
Then, for the intermediate mesons, the integral over $l^0$ in Eq.~\eqref{eq:two_point_0} changes into a sum over a set of 
$i\omega_n$,
\begin{align}
    &I_2(E,m_1,m_2,T)\notag\\
    =&-\int\frac{d^3  l}{(2\pi)^3}\frac{1}{\beta}
    \sum_n\frac{1}{(z_n^2-w_1^2+i\epsilon)[(E-z_n)^2-w_2^2+i\epsilon]},
    \label{eq:two_point_ft1}
\end{align}
where $z_n$ denotes $i\omega_n$.
The above summation of series is called the Matsubara sum. The Matsubara sum can be performed by the Sommerfeld-Watson transformation, i.e., a technique for converting the sum of the series into a contour integral on the complex plane using Cauchy's residue theorem. See the details in Appendix~\ref{sec:Sommerfeld-Watson}. Using Eq.~(\ref{convertion1}), the Matsubara sum in Eq.~(\ref{eq:two_point_ft1})
is computed as
\begin{align}
&I_2(E,m_1,m_2,T)\notag\\
=&\int\frac{d^3  l}{(2\pi)^3}
\oint\frac{dz}{2\pi i}
\frac{1}{(z^2-w_1^2+i\epsilon)[(E-z)^2-w_2^2+i\epsilon]}\notag\\
&\times\frac{1}{2}\coth\bigg(\frac{1}{2}\beta z\bigg)\notag\\
   =&\int\frac{d^3 l}{2(2\pi)^3}
   \bigg[\frac{\coth(\beta (w_1-i\epsilon)/2)}{2w_1[(E-w_1)^2-w_2^2+i\epsilon]}\notag\\
   &+\frac{\coth(\beta (w_1-i\epsilon)/2)}{2w_1[(E+w_1)^2-w_2^2+i\epsilon]}\notag\\
   &+\frac{\coth(\beta (w_2-E-i\epsilon)/2)}{2w_2[(E-w_2)^2-w_1^2+i\epsilon]}\notag\\
   &+\frac{\coth(\beta (w_2+E-i\epsilon)/2)}{2w_2[(E+w_2)^2-w_1^2+i\epsilon]}\bigg].
    \label{eq:two_point_ft2}
\end{align}
In the above calculation, we use the facts that $z_n$ represents the poles of the function $\coth(\beta z/2)$ and that the four terms in the last line stem from the poles of $\left((z^2-w_1^2+i\epsilon)[(E-z)^2-w_2^2+i\epsilon]\right)^{-1}$ on the complex $z$-plane. The details of the calculation are shown in Appendix~\ref{sec:2pft}.
As for the treatment of 
 the initial energy $E$, by following the prescription in Ref.~\cite{Gao:2019idb}, we adopt the periodic condition by setting $E=i2\pi n/\beta$ ($n\in \bm{\mathrm{Z}}$)  
 for the argument of hyperbolic cotangent function.
 Then, $I_{2}(E,m_1,m_2, T)$ in Eq.~\eqref{eq:two_point_ft2} is further simplified as
\begin{align}
  &I_2(E,m_1,m_2, T)\notag\\
=&\int\frac{d^3 l}{2(2\pi)^3}
\bigg[\frac{\coth(\beta (w_1-i\epsilon)/2)}{2w_1[(E-w_1)^2-w_2^2+i\epsilon]}\notag\\
&+\frac{\coth(\beta (w_1-i\epsilon)/2)}{2w_1[(E+w_1)^2-w_2^2+i\epsilon]}\notag\\
&+\frac{\coth(\beta (w_2-i\epsilon)/2)}{2w_2[(E-w_2)^2-w_1^2+i\epsilon]}\notag\\
&+\frac{\coth(\beta (w_2-i\epsilon)/2)}{2w_2[(E+w_2)^2-w_1^2+i\epsilon]}\bigg]. 
 \label{eq:two_point_ft3}
\end{align}
In order to calculate the integration over $\vec{l}$, we perform the numerical calculation by introducing a monopole-type form factor $f(\vec{l})=\Lambda^2/(\Lambda^2+|\vec{l}|^2)$ \cite{Kamano:2014zba} with the cutoff parameter $\Lambda$.
This form factor is introduced to represent the effect of finite sizes of hadrons, and plays the role of removing the ultraviolet divergence in the momentum integral.

The investigated temperature is limited to be not larger than 150 MeV.
This condition stems from the understanding that hadrons will be dissociated into quarks and gluons above about 150 MeV and hence the expressions by hadronic degrees of freedom cannot be used anymore. As the temperature goes to zero ($T\to0$, i.e., $\beta\to+\infty$), we notice that $I_2(E,m_1,m_2, T)$ in Eq.~(\ref{eq:two_point_ft3})
reproduces the expression of the scalar one-loop two-point function in vacuum given by
\begin{widetext}
\begin{align}
   I_{2,\text{vac}}(E,m_1,m_2)=&~\int\frac{d^3  lf(\vec{l})^2}{(2\pi)^3}\left(\frac{1}{2w_1[(E+w_1)^2-w_2^2+i\epsilon]}+\frac{1}{2w_2[(E-w_2)^2-w_1^2+i\epsilon]}\right)\notag\\
   =&~\int\frac{d^3 lf(\vec{l})^2}{(2\pi)^3}\frac{w_1+w_2}{2w_1w_2(E-w_1-w_2+i\epsilon)(E+w_1+w_2-i\epsilon)}.
   \label{eq:two_point_vac}
\end{align}
\end{widetext}
When the masses of two intermediate particles are the same, i.e., $m_1 = m_2$, Eq.~(\ref{eq:two_point_ft3})
becomes simple as
\begin{align}
    I_2(E,m,T)=\int\frac{d^3 lf(\vec{l})^2}{(2\pi)^3}\frac{\coth(\beta(w-i\epsilon)/2)}{w(E-2w)(E+2w)}.
    \label{eq:two_point_ft4}
\end{align}
\begin{figure}[h]
    \centering
    \includegraphics[width=0.9\linewidth]{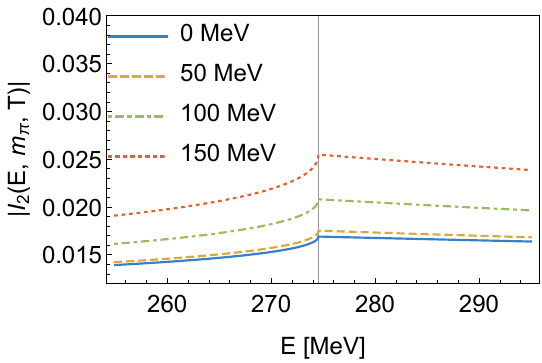}
    \caption{
    The temperature effect on the one-loop two-point function in the pion system in the isospin-symmetric case. The mass of the pion is given by the isospin-averaged mass $m_\pi$. The blue solid, yellow dashed, green dot-dashed, and red dotted lines are for the lineshape of $|I_2(E,m_\pi,T)|$ at temperatures $T=0$ MeV, $50$ MeV, $100$ MeV and $150$ MeV, respectively. The thin black solid vertical line marks the threshold of $\pi\pi$.}
    \label{fig:g2tft_12}
\end{figure}
From the above integrals, it is understood qualitatively that there exists an enhancement in the behaviors of the two-point function at finite temperature as the temperature $T$ increases.
This is confirmed by noting that the hyperbolic cotangent function $\coth(\beta(w_i-i\epsilon)/2)$ in the integral is the only temperature-related term in Eqs.~(\ref{eq:two_point_ft3}) and (\ref{eq:two_point_ft4}), and that the value of hyperbolic cotangent function quickly approaches one from infinity as its argument increases from zero to infinity. In this sense, the mass of the intermediate particle should not be too large in comparison to the temperature $T$ in order for that the prominent temperature effect on the threshold cusp appears. Therefore, we concentrate on a light meson system, i.e., pion system, to see the temperature effect on the one-loop two-point function.

Let us first consider the $\pi\pi$ system without isospin symmetry breaking by setting $m_{1}=m_{2}=m_{\pi}$ ($m_\pi$ is the isospin-averaged pion mass, i.e., $(2m_{\pi^\pm}+m_{\pi^0})/3$ \cite{pdg:2024ca}) as done in Eq.~\eqref{eq:two_point_ft4}.
We note that the temperature effects on the mass of pion are expected to be so small and negligible due to its Nambu-Goldstone boson nature \cite{Hatsuda:1986gu,Schenk:1993ca,Nicola:2014ca,Cheng:2011ca,Gu:2018swy}. So we suppose that the masses of pions does not change in temperature region $T\le 150~\mathrm{MeV}$.  This is expected because the masses of pions are protected by chiral symmetry as the Nambu-Goldstone bosons. 

By adopting Eq.~\eqref{eq:two_point_ft4}, the numerical results for $|I_2(E,m_\pi,T)|$ at different temperatures $T=0$ MeV, $50$ MeV, $100$ MeV, and $150$ MeV, are shown 
in Fig.~\ref{fig:g2tft_12}.
Notice that $I_{2,\text{vac}}$ in Eq.~(\ref{eq:two_point_vac}) in pion case without isospin symmetry breaking can produce exactly the same lineshape as shown by the blue line in Fig.~\ref{fig:g2tft_12} as a matter of fact. 
For the finite temperature effect, thus, it is demonstrated clearly in Fig.~\ref{fig:g2tft_12}
that the two-body threshold cusp is enhanced when the temperature $T$ increases.

The above simple analysis motivates us to study the $S$-wave $\pi\pi$ scattering process to investigate the cusp effect at finite temperature.
To be more realistic, however, we need to consider the isospin symmetry breaking: the mass difference between a neutral pion ($\pi^{0}$) and a charged pion ($\pi^{\pm}$).
This is the topic which is discussed in details in the next section.

\section{\boldmath Threshold cusp in $\pi\pi$ scattering at finite temperature}\label{sec:application1}

\subsection{\boldmath Lagrangian for $\pi\pi$ scattering}
\label{eq:pi_Lagrangian}
We consider the $S$-wave $\pi\pi$ scattering process by $\pi^{0}$ and $\pi^{\pm}$ at finite temperature.
For this purpose, we introduce the interaction chiral Lagrangian consisting of the chirally symmetric part $\mathcal{L}_S$ and the chirally broken part $\mathcal{L}_{SB}$ at the lowest order, i.e., 
\begin{align}  \mathcal{L}=&~\mathcal{L}_{S}+\mathcal{L}_{SB}\notag\\
   =&~\frac{1}{4}f_\pi^2
   \mathrm{tr}\!
   \left[(\partial_\mu U)(\partial^\mu U^\dagger)\right]\notag\\
   &+\frac{1}{2}\frac{m_\pi^2}{m_u+m_d}f_\pi^2
   \mathrm{tr}\!
   \left[
   M_l
   (U+U^\dagger-2)\right],
\end{align}
with the nonlinear representation of pion fields $U=e^{i\phi/f_\pi}$ and
\begin{align}
\phi
=\bm{\tau}\!\cdot\!\bm{\pi}
=\begin{pmatrix}
    \pi^0&\sqrt{2}\pi^+\\
    \sqrt{2}\pi^-&-\pi^0
\end{pmatrix}.
\end{align}
$M_l=
\mathrm{diag}
(m_u,m_d)$ is the diagonal matrix with up and down quark masses $m_u$ and $m_d$, respectively.
$f_\pi$ is the pion decay constant, $f_\pi=92.4~\mathrm{MeV}$, and $m_{\pi}$ is the isospin-averaged pion mass.
This Lagrangian provides us with the interactions between two pions in the particle base as
\begin{align}
    V_{\pi^0\pi^0\to\pi^0\pi^0} =&-\frac{m_\pi^2}{f_\pi^2},\\
    V_{\pi^0\pi^0\to\pi^+\pi^-} =&-\frac{1}{3f_\pi^2}(2s-t-u+m_\pi^2),\\
    V_{\pi^+\pi^-\to\pi^+\pi^-} =&-\frac{1}{3f_\pi^2}(s+t-2u+2m_\pi^2),\label{eq_chiral_interaction}
\end{align}
in correspondence to three 
scattering processes for neutral and charged pions: $\pi^0\pi^0\to\pi^0\pi^0$, $\pi^0\pi^0\to\pi^+\pi^-$ and $\pi^+\pi^-\to\pi^+\pi^-$.
Here $s$, $t$ and $u$ are the Mandelstam variables. We set $s=(p_1+p_2)^2$, $t=(p_1-p_3)^2$ and $u=(p_1-p_4)^2$, where $p_1$ and $p_2$ are the four-momenta of two incoming particles, and $p_3$ and $p_4$ are the four-momenta of two outgoing particles.
It is shown that, after partial wave analysis, the $S$-wave potential on each vertex is given by
\begin{align}
    V_{\pi}
  =&\begin{pmatrix}
        V_{\pi^{0}\pi^{0}\to\pi^{0}\pi^{0}} &
        V_{\pi^{0}\pi^{0}\to\pi^{+}\pi^{-}} \\
        V_{\pi^{+}\pi^{-}\to\pi^{0}\pi^{0}} &
        V_{\pi^{+}\pi^{-}\to\pi^{+}\pi^{-}}
    \end{pmatrix}\notag\\
  =&\begin{pmatrix}
        -\dfrac{m_\pi^2}{f_\pi^2}&-\dfrac{E^2-m_\pi^2}{f_\pi^2}\\[8pt]
        -\dfrac{E^2-m_\pi^2}{f_\pi^2}&-\dfrac{E^2}{2f_\pi^2}
    \end{pmatrix}.
    \label{eq:pion_kernel}
\end{align}

The propagator at finite temperature is written as
\begin{align}
    G_{\pi}
  =&\begin{pmatrix}
        G_{\pi^{0}\pi^{0}}&0\\
        0&G_{\pi^{+}\pi^{-}}
    \end{pmatrix}\notag\\
  =&\begin{pmatrix}
        \dfrac{1}{2}I_2(E,m_{\pi^0},T)&0\\[8pt]
        0&\dfrac{1}{2}I_2(E,m_{\pi^{\pm}},T)
    \end{pmatrix},
  \label{eq:pion_propagator}
\end{align}
including the neutral and charged pions in a matrix form
according to Eq.~(\ref{eq:two_point_ft4}).
The factor $1/2$ in the last equation accounts for the symmetry of identical particles.

\subsection{\boldmath Threshold cusp behavior in $\pi\pi$ scattering at finite temperature}
\label{sec:pipi_production}
The scattering $T$-matrix in the full scattering processes, denoted by $T$, is given by the Lippmann-Schwinger equation (LSE): $T=V+VGT$ with the interaction kernel $V$ and the propagator $G$. The $T$-matrix is solved as $T=(1-VG)^{-1}V$.
Let us consider generally the two channel case. The $T$, $V$ and $G$ are expressed by
\begin{align}
    T
  =\begin{pmatrix}
        T_{11} & T_{12} \\
        T_{21} & T_{22}
    \end{pmatrix},
    V
  =\begin{pmatrix}
        V_{11} & V_{12} \\
        V_{21} & V_{22}
    \end{pmatrix},
    G
  =\begin{pmatrix}
        G_{11} & 0 \\
        0 & G_{22}
    \end{pmatrix},
    \label{eq:TVG}
\end{align}
where the subscripts $i,j$ ($i,j=1,2$)
label the relevant channels. Solving the LSE, we find that the components in $T$ are given by
\begin{align}
    &T_{11}=\frac{V_{11}(1-G_{22}V_{22})+V_{12}G_{22}V_{21}}{(1-G_{11}V_{11})(1-G_{22}V_{22})-G_{11}V_{12}G_{22}V_{21}},\label{eq:T_matrix_general1}\\
    &T_{12}=T_{21}=\frac{V_{11}G_{11}V_{12}+V_{12}(1-G_{11}V_{11})}{(1-G_{11}V_{11})(1-G_{22}V_{22})-G_{11}V_{12}G_{22}V_{21}},\label{eq:T_matrix_general2}\\
    &T_{22}=\frac{V_{22}(1-G_{11}V_{11})+V_{21}G_{11}V_{12}}{(1-G_{11}V_{11})(1-G_{22}V_{22})-G_{11}V_{12}G_{22}V_{21}}.
    \label{eq:T_matrix_general3}
\end{align}
\begin{figure}[t]
    \centering
    \subfigure[The temperature effect on $T_{\pi^0\pi^0\to\pi^0\pi^0}$.]{
    \includegraphics[width=0.8\linewidth]{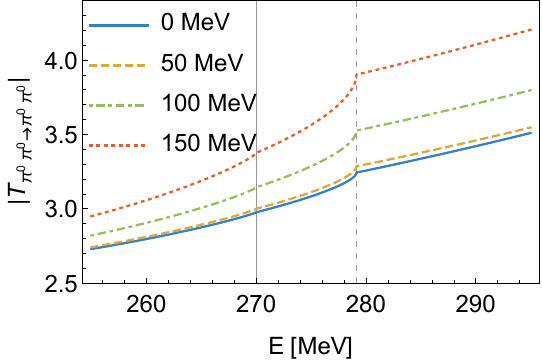}}
    \subfigure[The temperature effect on $T_{\pi^0\pi^0\to\pi^+\pi^-}$.]{
    \includegraphics[width=0.8\linewidth]{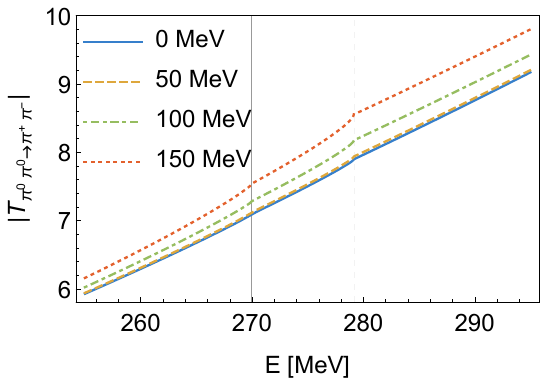}}
    \subfigure[The temperature effect on $T_{\pi^+\pi^-\to\pi^+\pi^-}$.]{
    \includegraphics[width=0.8\linewidth]{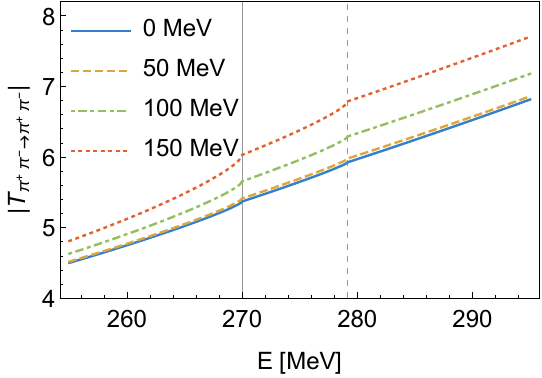}}
    \caption{
    The temperature effect on each element of $T_\pi$ in Eq.~\eqref{eq:T_pi}. The blue solid, yellow dashed, green dot-dashed, and red dotted lines are for the lineshape of the results at temperatures $T=0$ MeV, 50 MeV, 100 MeV, and 150 MeV, respectively. The black solid vertical line and black dashed vertical line indicate the thresholds of $\pi^0\pi^0$ and $\pi^+\pi^-$, respectively.}
    \label{fig:t123}
\end{figure} 
\begin{figure}[t]
    \centering
    \includegraphics[width=1\linewidth]{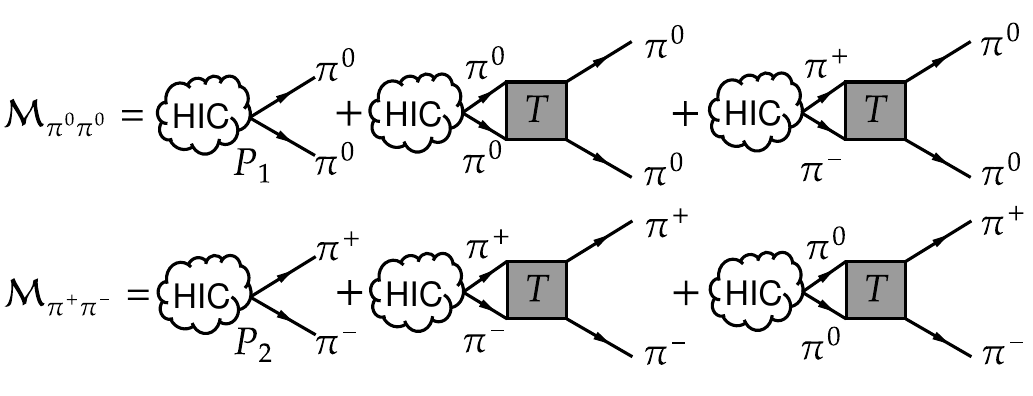}
    \caption{The complete production processes for different final states, $\pi^0\pi^0$ (upper) and $\pi^+\pi^-$ (lower). HIC is the abbreviation of Heavy Ion Collision.}
    \label{fig:pamplitude1}
\end{figure}
We apply this formula to the $\pi\pi$ scattering by substituting the scattering kernel in Eq.~\eqref{eq:pion_kernel} and propagator in Eq.~\eqref{eq:pion_propagator} into $V$ and $G$, respectively.
Then, we can get 
\begin{align}
    T_{\pi}
  =\begin{pmatrix}
        T_{\pi^{0}\pi^{0} \rightarrow \pi^{0}\pi^{0}} & T_{\pi^{0}\pi^{0} \rightarrow \pi^{+}\pi^{-}} \\
        T_{\pi^{+}\pi^{-} \rightarrow \pi^{0}\pi^{0}} & T_{\pi^{+}\pi^{-} \rightarrow \pi^{+}\pi^{-}}
    \end{pmatrix},
    \label{eq:T_pi}
\end{align}
and investigate the temperature effect on $T_\pi$ for $\pi\pi$ scatterings including isospin symmetry breaking.
The numerical results are displayed in Fig.~\ref{fig:t123}.
We find that the two-body threshold cusps at two thresholds become more visible with the increasing temperature $T$.
It should be noted that the temperature dependence on the pion mass and decay constant are not considered in our calculation.
This is because the change of pion mass is less dependent on the temperature as shown 
in Ref.~\cite{Gu:2018swy} (see Fig.~7 in the reference).
Furthermore, we think that the temperature dependence of pion decay constant on the temperature would be less significant according to the pion mass dependence of the pion decay constant as shown in Ref.~\cite{Gu:2018swy} (see Fig.~5 in the reference). Thus, we can neglect the temperature effect on the pion mass and pion decay constant simultaneously.

\section{Production in the environment of finite temperature.}\label{Production}
\subsection{ Production rate}
Now we consider a realistic situation to see the finite temperature effect on the scattering amplitude $T$, 
which can be, for example, in heavy ion collisions.  
The relevant amplitude is written by an initial production process $P_{i}$ and the subsequent scattering process via $T$ 
(final state interaction). According to Ref.~\cite{Zhuang:2021pci}, the total amplitude of particle production is given by
\begin{align}
   \mathcal{M}=P+PGT,
   \label{eq:tot_amp}
\end{align}
which includes the bare (direct) vertex for production process
and the iterated vertex for scattering process, as shown in the first and second terms, respectively, in the right-hand side.
The bare production amplitude
\begin{align}
    P=(P_1,P_2),
    \label{eq:P1_P2}
\end{align}
represents the direct production processes of the channel 1 and 2, respectively. 

 In heavy-ion collisions, the meson pairs $\pi^0\pi^0$ and $\pi^+\pi^-$ could be produced by, for instance,  the other hadronic decays through weak interactions, the direct hadronization processes and so on. Therefore, $P_i$ can take arbitrary values in general, and its magnitude is relevant to the bare production rate, which could modify the temperature effect on the cusp structure. 
However, there are several reasons that we expect that the cusp behavior is not influenced by the different values of $P_{1,2}$.  
First, we note that the threshold cusp is a kinematical effect which originates from a square root branch point in the complex plane of energy because of the on-shellness of the intermediate particles in the scattering process.  
Second, it is reasonable to expect that the  production mechanism is dominated by the strong interaction, and therefore, the strengths $P_{1,2}$ are expected to be similar. 
Nevertheless, in the case that $P_{1,2}$ are different, we have checked explicitly that the cusp behavior remains qualitatively the same with different $P_{1,2}$ values.  
The last point is the effect of the relative phase of $P_{1,2}$, which is not determined.  
To show this influence we have performed calculations for the two extreme cases where $P_{1,2}$ take opposite signs with the same strength, 
the results of which will be discussed carefully later in this section.

Now from  Eq.~\eqref{eq:tot_amp}, as a result, we obtain the general form of the total production
\begin{align}
&\mathcal{M}_{1}=P_1+P_1G_{11}T_{11}+P_2G_{22}T_{21},
\label{eq:total_amplitude_production_1} \\
&\mathcal{M}_{2}=P_2+P_2G_{22}T_{22}+P_1G_{11}T_{12},
\label{eq:total_amplitude_production_2}
\end{align}
where $\mathcal{M}_{i}$ ($i=1,2$) is the component of the total amplitude $\mathcal{M}$.

Applying the general expressions~\eqref{eq:total_amplitude_production_1} and \eqref{eq:total_amplitude_production_2} for $\pi^{0}\pi^{0}$ and $\pi^{+}\pi^{-}$ cases, respectively, as depicted by Fig.~\ref{fig:pamplitude1}, we obtain the total amplitudes ${\cal M}_{\pi}$ from the heavy-ion collisions.
It has two components, i.e., ${\cal M}_{\pi^0\pi^0}$ and ${\cal M}_{\pi^+\pi^-}$ for the $\pi^{0}\pi^{0}$ and $\pi^{+}\pi^{-}$ productions, respectively, which are given by
\begin{align}
\mathcal{M}_{\pi^0\pi^0}
=&P_{\pi^{0}\pi^{0}}
+P_{\pi^{0}\pi^{0}}G_{\pi^{0}\pi^{0}}T_{\pi^{0}\pi^{0} \rightarrow \pi^{0}\pi^{0}}
\notag\\
&+P_{\pi^{+}\pi^{-}}G_{\pi^{+}\pi^{-}}T_{\pi^{+}\pi^{-} \rightarrow \pi^{0}\pi^{0}},
\label{eq:pi_production_amplitude_1} \\
\mathcal{M}_{\pi^+\pi^-}
=&P_{\pi^{+}\pi^{-}}
+P_{\pi^{+}\pi^{-}}G_{\pi^{+}\pi^{-}}T_{\pi^{+}\pi^{-} \rightarrow \pi^{+}\pi^{-}}\notag\\
&+P_{\pi^{0}\pi^{0}}G_{\pi^{0}\pi^{0}}T_{\pi^{0}\pi^{0} \rightarrow \pi^{+}\pi^{-}}.
\label{eq:pi_production_amplitude_2}
\end{align}
\begin{figure*}[t]
    \centering
    \subfigure[$P_{\pi^0\pi^0}=1$, $P_{\pi^+\pi^-}=1$]{
   \includegraphics[width=0.45\linewidth]{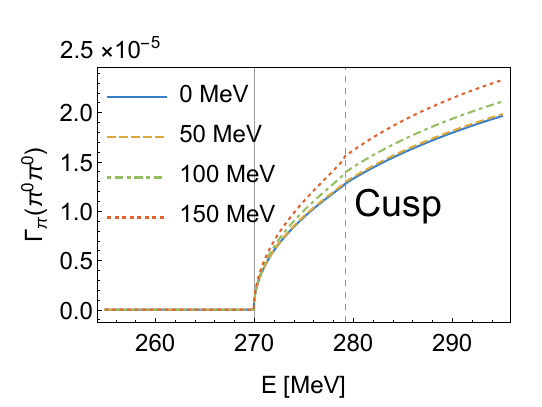}
   }
    \subfigure[$P_{\pi^0\pi^0}=1$, $P_{\pi^+\pi^-}=-1$]{
 \includegraphics[width=0.45\linewidth]{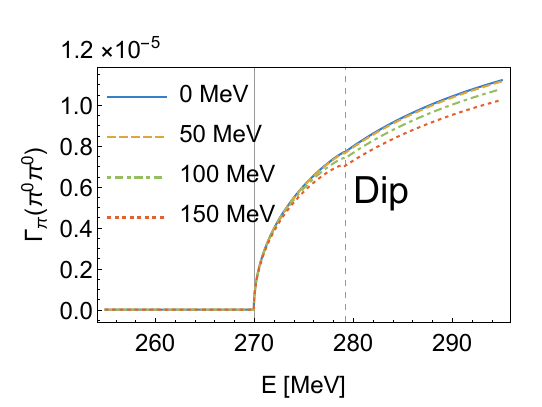}
 }
    \subfigure[$P_{\pi^0\pi^0}=1$, $P_{\pi^+\pi^-}=1$]{
  \includegraphics[width=0.45\linewidth]{ 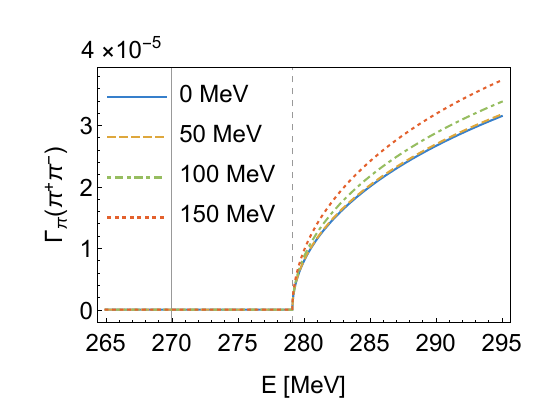}
  }
    \subfigure[$P_{\pi^0\pi^0}=1$, $P_{\pi^+\pi^-}=-1$]{   \includegraphics[width=0.45\linewidth]{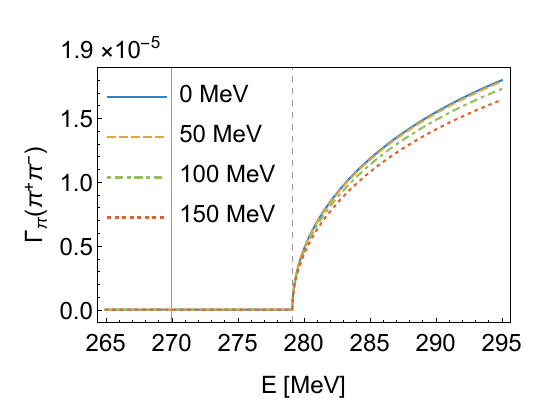}}
    \caption{
    The temperature effect on the $\pi\pi$ production rate $\Gamma_{\pi}$ in Eq.~\eqref{eq:prod_amp_pion}: the $\pi^0\pi^0$ channels in the panels (a) and (b) and the $\pi^+\pi^-$ channels in the panels (c) and (d).
   The cases of the same and opposite relative signs in $(P_{\pi^{0}\pi^{0}},P_{\pi^{+}\pi^{-}})$ are shown.
    The blue solid, yellow dashed, green dot-dashed, and red dotted lines are for the lineshape of the results at temperatures $T=0$ MeV, 50 MeV, 100 MeV, and 150 MeV, respectively. The solid and dashed black vertical lines indicate the thresholds of $\pi^0\pi^0$ and $\pi^+\pi^-$, respectively.
    See the text for more details.}
    \label{fig:pamplitude2}
\end{figure*}

We remember that the invariant mass distribution of two particles in the rest frame of the parent particle 
can be obtained by
\begin{equation}
   d\Gamma = \frac{(2\pi)^4}{2M}|\mathcal{M}|^2
   d\Phi_2(\vec{p}_1,\vec{p}_2,\vec{p}_3),
   \label{eq:inv_mass_distribution}
\end{equation}
with $M$ and $\vec{p}_1$ being the mass and the three momentum of the parent particle, $\vec{p}_2$ and $\vec{p}_3$ the three momenta of two decaying particles, respectively, in the final states in the decay. Here $\Phi_2(\vec{p}_1,\vec{p}_2,\vec{p}_3)$ is the two-body phase-space volume of the decay process with two decaying particles. Substituting the pion scattering amplitude ${\cal M}_{\pi}$ into Eq.~\eqref{eq:inv_mass_distribution} and performing the momentum integrals, we obtain the production rate of two pions as
\begin{equation}
     \Gamma_{\pi}
     =
     \frac{1}{8\pi M^2}|\vec{p}||\mathcal{M}_{\pi}|^2
     =
     \frac{1}{8\pi E^2}\sqrt{\frac{1}{4}E^2-m_\pi^2}|\mathcal{M}_{\pi}|^2,
     \label{eq:prod_amp_pion}
\end{equation}
in the rest frame of the initial state,
where $\vec{p}$ (or $-\vec{p}$) is the three momentum of the outgoing particle in the rest frame and $E$ is the total energy.

Notice that
the phase space 
for $\pi^0\pi^0$ scattering is multiplied by
an additional factor $1/2$ since two $\pi^0$s are identical particles.
It is also important to recognize that the production rate $\Gamma_{\pi}$ in Eq.~\eqref{eq:prod_amp_pion} is dependent, not only on the total energy $E$, but also on the temperature $T$ through the production amplitude $\mathcal{M}_{\pi}$.

We show the temperature effect on the production rate $\Gamma_{\pi}$
in Fig.~\ref{fig:pamplitude2}.
In the figure, the results are shown for the different combinations of the relative signs in $(P_{1},P_{2})=(P_{\pi^{0}\pi^{0}},P_{\pi^{+}\pi^{-}})$:
(a) $(1,1)$ and (b) $(1,-1)$ for $\pi^{0}\pi^{0}$ productions in Eq.~\eqref{eq:pi_production_amplitude_1} and (c) $(1,1)$ and (d) $(1,-1)$ for $\pi^{+}\pi^{-}$ productions in Eq.~\eqref{eq:pi_production_amplitude_2}.
We observe several interesting features.

Firstly, as shown in Fig.~\ref{fig:pamplitude2}(a) and (b), the $\pi^{0}\pi^{0}$ production rates have a cusp or dip structure at the $\pi^{+}\pi^{-}$ threshold depending on the relative sign of $(P_{\pi^{0}\pi^{0}},P_{\pi^{+}\pi^{-}})$.
There is a {\it cusp} in the case of the same relative sign, i.e., $(P_{\pi^{0}\pi^{0}},P_{\pi^{+}\pi^{-}})=(1,1)$, while there is a {\it dip} in case of the opposite relative sign, i.e., $(P_{\pi^{0}\pi^{0}},P_{\pi^{+}\pi^{-}})=(1,-1)$.
Such a difference may be small in the zero temperature (vacuum) case, and the experimental observations may be difficult.
In contrast, it is enhanced as the temperature increases, and hence observing the $\pi^{0}\pi^{0}$ production rate in HIC gives us clear information on the $\pi\pi$ scatterings.

Secondly, the temperature effect behaves differently for the same and opposite relative signs in $(P_{\pi^{0}\pi^{0}},P_{\pi^{+}\pi^{-}})$, as seen commonly in both $\pi^{0}\pi^{0}$ and $\pi^{+}\pi^{-}$ productions. This is due to the interference between the direct and rescattering processes, in contrast to the ordinary S-wave scatterings where the threshold behavior appears as either a cusp or (reflected) S-like shape structure. 
In the case of the same relative sign, i.e., $(P_{\pi^{0}\pi^{0}},P_{\pi^{+}\pi^{-}})=(1,1)$, the finite temperature {\it enhances} the magnitudes of the production rates.
In contrast, in the case of the opposite relative sign, i.e., $(P_{\pi^{0}\pi^{0}},P_{\pi^{+}\pi^{-}})=(1,-1)$, the finite temperature {\it suppresses} the magnitudes of the production rates.
Thus, the relative signs in $(P_{\pi^{0}\pi^{0}},P_{\pi^{+}\pi^{-}})$ are intimately related to the behaviors of the production magnitudes at finite temperatures.

As the results in Fig.~\ref{fig:pamplitude2}, we summarize that the finite temperature effect acts as a magnifying glass for studying the $\pi\pi$ scattering: the existence of the cusp or dip structure and the enhancement or suppression of the magnitude of the production rate.
Therefore, we conclude that experimental observations of the production rate in HIC provide us with the important information on $\pi\pi$ productions.

 \begin{figure}[t]
    \centering
    \includegraphics[width=0.9\linewidth]{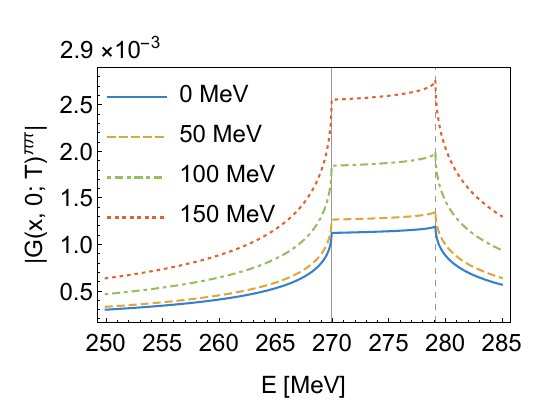}
    \caption{The temperature effect on the absolute value  of $\pi\pi$ propagator, $|G(x,0;T)^{\pi\pi}|$, where the isospin symmetry breaking is shown, see Eq.~\eqref{eq:pipi_propagator}.
    The blue solid, yellow dashed, green dot-dashed, and red dotted lines are for the lineshape of the results at temperatures $T=0$ MeV, 50 MeV, 100 MeV, and 150 MeV, respectively.
    The solid and dashed black vertical lines indicate the thresholds of $\pi^0\pi^0$ and $\pi^+\pi^-$, respectively.
    The plateau-like structures are seen between $\pi^{0}\pi^{0}$ and $\pi^{+}\pi^{-}$ thresholds for each temperature.}
    \label{fig:isobtem_pion}
\end{figure}

\subsection{\boldmath  Propagator and isospin breaking}
\label{sec:pipi_propagator_plataeu}

 The temperature effect on the threshold cusp can also be seen clearly in the $\pi\pi$ propagator which includes the isospin symmetry breaking. The $\pi\pi$ propagator from the position $0$ to $x$ is defined by
 \begin{align}
    &G(x,0;T)^{\pi\pi}=\langle0|[\pi(x)\pi(x)]_{I=0}[\pi(0)\pi(0)]_{I=2}|0\rangle\notag\\
    =&\frac{\sqrt{2}}{3}[I_2(E,m_{\pi^+},m_{\pi^-},T)-I_2(E,m_{\pi^0},m_{\pi^0},T)],
   \label{eq:pipi_propagator}
\end{align}
 in the propagating process which connects the total isospin $I=2$ and $I=0$ states.
Here $I_2(E,m_{\pi^+},m_{\pi^-},T)$ and $I_2(E,m_{\pi^0},m_{\pi^0},T)$ are the two-point functions at finite temperature $T$ for $\pi^+\pi^-$ and $\pi^0\pi^0$, respectively. See Appendix~\ref{sec:isobreaking} for the detailed calculation. If the isospin symmetry is exactly conserved, the $\pi\pi$ propagator $G(x,0;T)^{\pi\pi}$ vanishes. However, if the isospin symmetry is broken, then $G(x,0;T)^{\pi\pi}$
becomes nonzero: $G(x,0;T)^{\pi\pi}\neq0$.

Let us investigate the behaviors of $G(x,0;T)^{\pi\pi}$ at various temperatures in the isospin-breaking case. In Fig.~\ref{fig:isobtem_pion}, we show $|G(x,0;T)^{\pi\pi}|$ as functions of the total energy $E$ for several temperatures $T=0$ MeV, $50$ MeV, $100$ MeV, and $150$ MeV. There are plateau-like structures emerging between the $\pi^0\pi^0$ threshold and the $\pi^+\pi^-$ threshold. We note that the isospin symmetry breaking effect plays an important role to form the plateau-like structure, because such a plateau should shrink to become only a peak (or a cusp) when the isospin symmetry is recovered, i.e., the interval between the $\pi^{0}\pi^{0}$ threshold and the $\pi^{+}\pi^{-}$ threshold vanishes.
 The heights of the plateaus are significantly enhanced as the temperature increases.
 This
 should be one of the most important mechanisms for producing the enhancement or suppression of the $\pi\pi$ production rate, as shown in Fig.~\ref{fig:pamplitude2}.

 In the end of this section, we comment on the electromagnetic effect which contributes to the isospin breaking as well as the quark mass difference.   In Ref.~\cite{Knecht:1997jw} they have studied such effect in the  chiral perturbation theory and have shown that the electromagnetic effect is not very important for the scattering length.  Therefore, we expect that the effect on the threshold cusp is also not very important.  A very crude estimate of the electromagnetic interaction can be done by comparing the interaction strength of the Coulomb potential and that of the chiral interaction~(\ref{eq_chiral_interaction}).  For the former, we use the volume integrated potential in the volume of the twice (because of the interaction) of the Compton wave length scale of the pion.  It gives
\begin{align}
V_{\text{EM}} \sim \int_0^{2/m_\pi} d^3 x \frac{\alpha}{r} = \alpha \frac{8 \pi}{m_\pi^2} \sim 0.4\ {\rm fm}^2,
\label{eq:V_EM}
\end{align}
where $\alpha=1/137$ is the fine structure constant.
For the latter, the corresponding quantity is 
\begin{align}
V_{\pi \pi} \sim \frac{m_\pi^2}{f_\pi^2} \frac{1}{2 m_\pi^2} \sim 2\ {\rm fm}^2, \label{eq:V_pipi}
\end{align}
where the last factor is from the normalization of the boson wave function. The relevant EM interaction of Eq.~\eqref{eq:V_EM} is for the $\pi^+ \pi^-$ channel, which is added to
the strong interaction in Eq.~\eqref{eq:V_pipi} as an attractive interaction, denoted by $V_{\text{EM}}$.  
In Fig. 7, we compare $T^{\rm full}_{\pi^0 \pi^0 \to \pi^0 \pi^0}$ with and without $V_{\rm EM}$ at zero and finite temperature $T = 100$ MeV, i.e., $T=0.507$ fm$^{-1}$. It indicates that the electromagnetic effect is small and it does not influence the cusp structure too much as compared to the temperature effect. Therefore, to address the temperature effect on the cusp, we can safely ignore the electromagnetic effect. 
\begin{figure}
    \centering
    \includegraphics[width=0.9\linewidth]{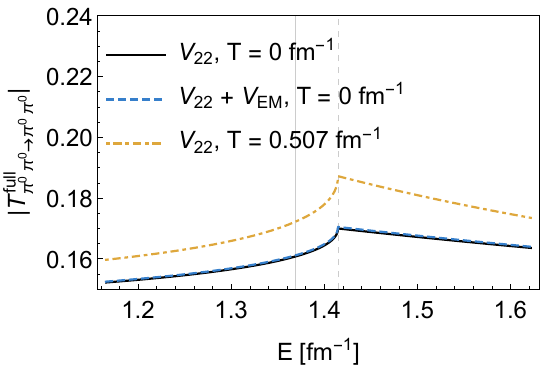}
    \caption{A rough estimation of the contribution from $V_{\text{EM}}$ to the $|T^{\text{full}}_{\pi^0\pi^0\to\pi^0\pi^0}|$. $V_{11}=-0.1$~fm$^2$, $V_{12}=V_{21}=-2.0$~fm$^2$, $V_{22}=-2.0$~fm$^2$.}
    \label{fig:EM_effect}
\end{figure}

\section{\boldmath Threshold cusp in $D\bar{D}^\ast$ scattering at finite temperature}
\label{sec:application2}

\begin{figure}[t]
    \centering
    \subfigure[The temperature dependence of the
    $D$ and $\bar{D}^{\ast}$ meson
    masses.]{
    \includegraphics[width=0.9\linewidth]{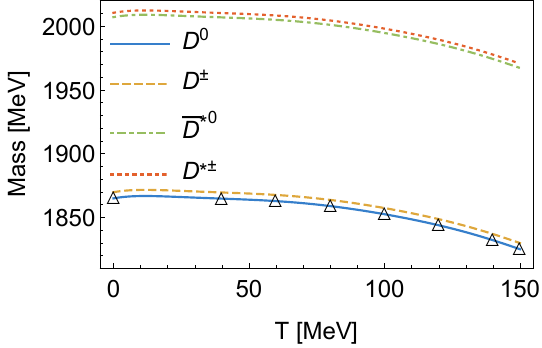}}
    \subfigure[The temperature dependence of the
    $D$ and $\bar{D}^{\ast}$ decay widths.]{
    \includegraphics[width=0.9\linewidth]{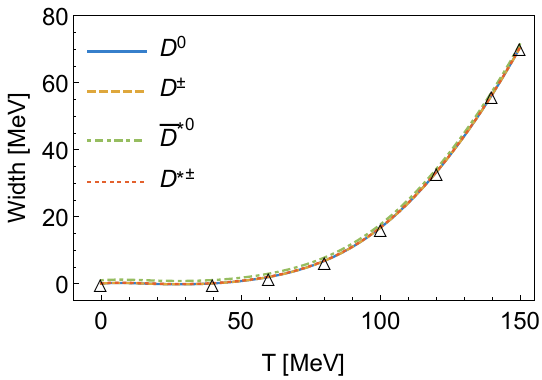}}
    \caption{The temperature dependence of the
    $D$ and $\bar{D}^{\ast}$ meson
    masses (a) and decay widths (b). The black triangles denote the theoretical values of a $D$ meson extracted from Ref.~\cite{Montana:2020lfi}. }
    \label{fig:Dmass_width_tem}
\end{figure}

The formalism in Sec.~\ref{sec:two_point}
can be adopted to
study the temperature effect on the threshold cusp in charmed meson scatterings.
We apply the similar discussion to the process in $D\bar{D}^{\ast}$ (or $D^{\ast}\bar{D}$) meson system by including the isospin symmetry breaking.
The $D\bar{D}^{\ast}$ (or $D^{\ast}\bar{D}$) meson system is important because it is intimately related to the exotic hadrons like $X(3872)$. Here we focus on the $D\bar{D}^{\ast}$ channel for a simple discussion, because excluding the $D^{\ast}\bar{D}$ channel does not change our conclusion. We introduce the $D\bar{D}^{\ast}$ meson propagator including the modified masses and decay widths in hot medium. The modifications of the masses and decay widths are parameterized based on the results obtained by the other studies of hadron model~\cite{Montana:2020lfi,Fuchs:2004fh,Cleven:2017fun}.

With this setup, we consider the $D\bar{D}^{\ast}$ meson propagator connecting the total isospin $I=1$ and $I=0$ states, given by
\begin{align}
    &G(x,0;T)^{D\bar{D}^\ast}\notag\\
    =&\langle0|[D(x)\bar{D}^*(x)]_{I=0}[D(0)\bar{D}^*(0)]_{I=1}|0\rangle\notag\\
    =&
    \frac{1}{2}
    I_2\biggl(E,m_{D^0}(T)-\frac{1}{2}i\Gamma_{D^0}(T),m_{\bar{D}^{*0}}(T)-\frac{1}{2}i\Gamma_{\bar{D}^{*0}}(T),T\biggr)\notag\\
    &-
    \frac{1}{2}
    I_2\biggl(E,m_{D^+}(T)-\frac{1}{2}i\Gamma_{D^+}(T),m_{D^{*-}}(T)\notag\\
    &-\frac{1}{2}i\Gamma_{D^{*-}}(T),T\biggr),
    \label{eq:isobDDs}
\end{align}
where
the former and latter $I_{2}$ functions
are the two-point functions at finite temperature $T$ for $D^0\bar{D}^{*0}$ and $D^+D^{*-}$, respectively. See Appendix~\ref{sec:isobreaking} for the detailed calculation. The temperature dependence of the masses and decay widths of $D$ and $\bar{D}^{\ast}$ mesons are parameterized by $m(T)$ and $\Gamma(T)$, and their function forms are extracted by polynomial functions from Ref.~\cite{Montana:2020lfi}, as shown in Fig.~\ref{fig:Dmass_width_tem}.

\begin{figure}[t]
    \centering
    \includegraphics[width=0.9\linewidth]{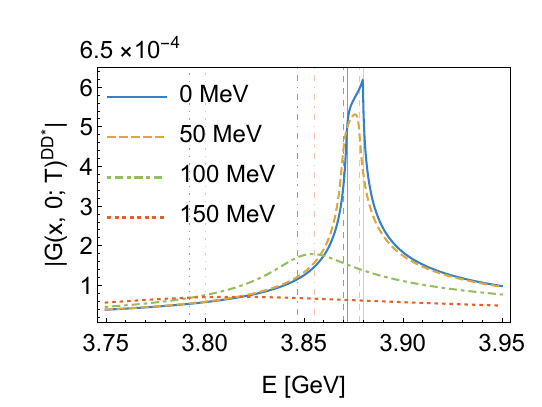}
    \caption{The temperature effect on the absolute value of $D\bar{D}^*$ propagator, $|G(x,0;T)^{D\bar{D}^\ast}|$, where the isospin symmetry breaking is shown, see Eq.~(\ref{eq:isobDDs}). The blue solid, yellow dashed, green dot-dashed, and red dotted lines are for the lineshape of the results at temperatures $T=0~\mathrm{MeV}, ~50~\mathrm{MeV},~100~\mathrm{MeV}$, and 150 MeV, respectively. The black vertical lines and red vertical lines
    denote the mass thresholds of $D^0\bar{D}^{*0}$ and $D^+D^{*-}$, respectively.}
    \label{fig:isobtem_D}
\end{figure}  

In Fig.~\ref{fig:Dmass_width_tem},
the triangles in the panels (a) and (b) display the theoretical values for the masses and decay widths of a $D$ meson at different temperatures which were calculated in Ref.~\cite{Montana:2020lfi}.
They were obtained by studying the hadron model with chiral and heavy-quark spin-flavor
symmetries by adopting the imaginary time formalism.
As shown in Fig.~\ref{fig:Dmass_width_tem}(a), the mass of the $D$ meson decreases as the temperature increases. 
For example, the mass drops by around 40 MeV at temperature $T=150$ MeV. In Fig.~\ref{fig:Dmass_width_tem}(b), it is shown that the decay width of the $D$ meson increases as the temperature increases. These behaviors also agree qualitatively with the calculation result for the mass and width of the $D$ meson obtained in hot pion medium in Ref.~\cite{Fuchs:2004fh}, where the in-medium $D$ meson self-energies were estimated by the scattering length of the $D$ meson in a hot pion gas. Similar medium modifications were found for $D^*$ vector mesons \cite{Fuchs:2004fh}.
Apart from Ref.~\cite{Fuchs:2004fh}, it is shown in Ref.~\cite{Cleven:2017fun} that the medium effect on the decay width of a $D$ meson at finite temperature exhibits the similar behaviors to those of a $D^{\ast}$ meson. Therefore, in the present study, we assume that the mass drop and the increase of the decay width of a $\bar{D}^\ast$ meson behave similarly to those of the $D$ meson.
Thus, we apply the $m(T)$ and $\Gamma(T)$ function forms, estimated for a $D$ meson in Ref.~\cite{Montana:2020lfi}, to a $\bar{D}^\ast$ meson at finite temperature, as
shown in Fig.~\ref{fig:Dmass_width_tem}.

Similarly to 
the $\pi\pi$ propagator in Eq.~\eqref{eq:pipi_propagator},
we find that the $D\bar{D}^\ast$ propagator vanishes if the isospin symmetry is exactly conserved.
It is also the case that, if the isospin symmetry is broken, we obtain nonzero values, $G(x,0;T)^{D\bar{D}^\ast}\neq 0$, and
find plateau-like structures emerging between the mass thresholds of $D^0\bar{D}^{*0}$ and $D^+D^{*-}$ in the lineshape of $|G(x,0;T)^{D\bar{D}^\ast}|$, as shown in Fig.~\ref{fig:isobtem_D}.
The plateau-like structure is clearly seen at the zero temperature case.
At higher temperatures, however,
the plateau-like structures are qualitatively different from those in the $\pi\pi$ case, as summarized in the following.

Firstly,
the plateau-like structure is shifted to the lower energy region, since the masses of the
$D$ and $\bar{D}^{\ast}$ mesons decrease with
the increasing temperature as shown in Fig.~\ref{fig:Dmass_width_tem}(a).
The height of the plateau-like structure decreases as the temperature increases,
and eventually it disappears at $T=150~\mathrm{MeV}$, as shown in Fig.~\ref{fig:isobtem_D}. Secondly, the plateau-like structure becomes smooth as the temperature increases. In other words, the plateau-like structure is not seen clearly at higher temperatures. The disappearance of the the plateau-like structure of the $D\bar{D}^{\ast}$ case is sharply different from the $\pi\pi$ case. This property stems from that the decay widths of the $D$ and $\bar{D}^{\ast}$ mesons increase with the increasing temperature, as shown in Fig.~\ref{fig:Dmass_width_tem}(b).
As a summary, we conclude that the isospin symmetry breaking in $D\bar{D}^*$ scatterings from HIC brings us important information which is qualitatively different from the $\pi\pi$ case.

\section{Summary and outlook}\label{sec:sum}

We have
investigated the temperature effects on the threshold cusps in $\pi\pi$ and $D\bar{D}^\ast$ scatterings. The temperature is introduced in 
the one-loop two-point function by using the Sommerfeld-Watson transformation after introducing the Matsubara frequencies. 
 The temperature effect has been investigated below $150~\mathrm{MeV}$ 
 due to the applicability of the hadronic degrees of freedom.
 We have evaluated the production rate
 of pions
 at different temperatures,
 and studied how the temperature affects the threshold cusp in the $\pi\pi$ scatterings.
 From the lineshape of the production rate in $\pi\pi$ scatterings,
 we have shown that, as the temperature increases, the cusp around the $\pi^+\pi^-$ threshold becomes enhanced in the case of the bare production amplitude $(P_{\pi^0\pi^0},P_{\pi^+\pi^-})=(1,1)$,
 while the dip becomes suppressed in the case of $(P_{\pi^0\pi^0},P_{\pi^+\pi^-})=(1,-1)$.
 Thus, the temperature dependence of the production rate near the $\pi^{+}\pi^{-}$ threshold is very sensitive to the relative sign in the bare production amplitude.
 We also find that the lineshape of 
 the $\pi\pi$ propagator
  displays some unique properties at different temperatures: plateau-like structures appearing between the $\pi^0\pi^0$ and $\pi^+\pi^-$ thresholds due to isospin symmetry breaking.
  The heights of the plateau-like structures become much enhanced with the increasing temperature. For comparison, we also study the temperature effect on the threshold cusp in $D\bar{D}^\ast$ scatterings 
  including isospin symmetry breaking.
  The lineshape of 
  the $D\bar{D}^{\ast}$ propagator
  presents a plateau-like structure
  similar to the $\pi\pi$ case. However, it disappears as the temperature increases, because the masses and decay widths of $D$ and $\bar{D}^\ast$ mesons are 
  modified at finite temperature. 

The present formalism can be applied, not only to boson-boson systems, but also to boson-fermion and fermion-fermion systems.
Indeed, there are several candidates of cusp or dip structures reported in experiments: $\Sigma N$ ($I=1/2$) cusp~\cite{Dalitz:1980zc}, $\bar{K}N$ ($I=1$) cusp~\cite{Belle:2022ywa}, $\eta\Lambda$ cusp~\cite{Zhang:2024jby} and so on. As for $\Sigma N$ ($I=1/2$) and $\bar{K}N$ ($I=1$), we can introduce the isospin symmetry breaking and may be able to study the plateau-structure as we have discussed in the $\pi\pi$ scatterings.
The $\eta\Lambda$ cusp is relevant to the property of $\Lambda(1670)$.
As far as the hadron masses and decay widths are less modified in hot medium, we can expect that the cusp or dip structures around the thresholds become enhanced or suppressed and they bring us important information on the hadron scatterings.
Further studies are left for future.

\vspace{0.2cm}
{\bf \color{gray}Acknowledgements:}~~
We are grateful to Xingyu Guo, Pengyu Niu, Xinyue Hu, and Quanxing Ye for the helpful discussion. 
This work is partly supported by the National Natural Science Foundation of China with Grant Nos.~12375073 and ~12035007, Guangdong Provincial funding with Grant No.~2019QN01X172. A.~H. is supported in part by the JSPS Grants-in-Aid for Scientific Research with Grant Nos. 21H04478(A) and 24K07050(C). Y.~Z. is supported in part by the Guangdong Provincial international exchange program for outstanding young talents of scientific research in 2023.

\clearpage
\nocite{*}

\newpage
\appendix
\section{Summation of series: Sommerfeld-Watson transformation}\label{sec:Sommerfeld-Watson}
\begin{figure}[h]
    \centering
    \includegraphics[width=0.9\linewidth]{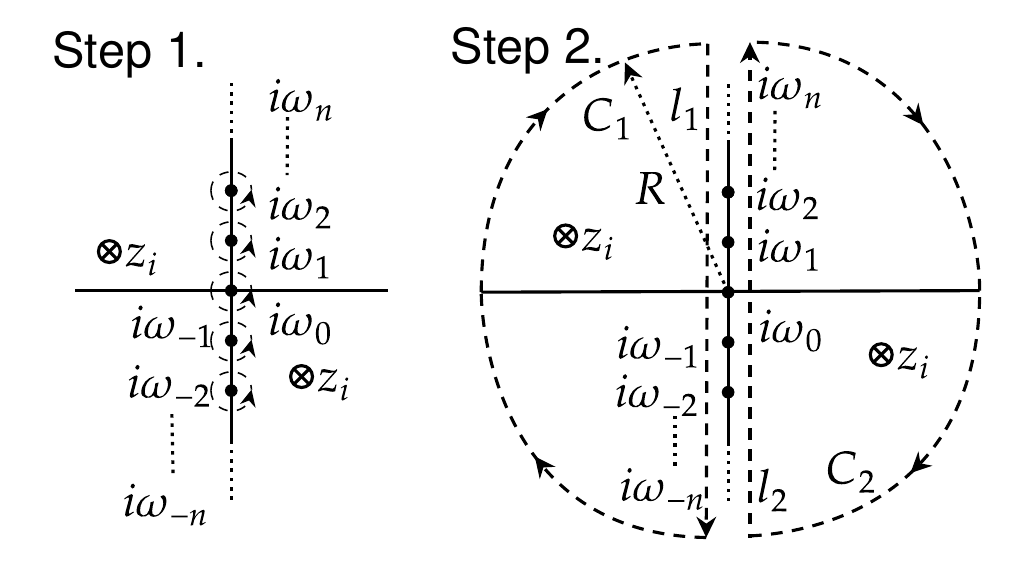}
    \caption{
    The transformation of the integration contour for Eq.~(\ref{matsubara_sum}).}
    \label{matsubara_sum}
\end{figure}
Assume that $f(z)$ is a function which is analytic at $z=i2\pi n/\beta$ $(n\in \bm{\mathrm{Z}})$, and tends to zero as fast as, or faster than
\begin{align}
    \frac{1}{|z|^2}~~\text{as}~~|z|\to\infty.
    \label{eq_condition}
\end{align}
 And consider the following sum: 
\begin{equation}
    S = \sum^{\infty}_{n=-\infty}f(n).
    \label{matsubara_sum}
\end{equation}
To compute this discrete sum, let us introduce the function $F(z)$ as
\begin{equation}
    F(z)\equiv f(z)\frac{1}{2}\coth\biggl(\frac{z}{2}\beta\biggr)=f(z)\frac{1}{2}\frac{e^{\beta z}+1}{e^{\beta z}-1}.
\end{equation}
This function has simple poles associated with the hyperbolic cotangent function at $z_n=i\omega_n=i2\pi n/\beta$ with residues
\begin{align}
    &\text{Res}(F(z),z=i\omega_n)\notag\\=&\lim_{z\to i\omega_n}f(z)\frac{1}{2}\frac{e^{\beta z}+1}{\frac{d}{dz}(e^{\beta z}-1)}\notag\\
    =&\lim_{z\to i\omega_n}f(z)\frac{1}{2}\frac{e^{\beta z}+1}{\beta e^{\beta z}}=\frac{1}{\beta}\sum_nf(z_n),
    \label{eq149}
\end{align}
and the poles associated with $f(z)$ at $z=z_i$.

First, we transform the sum of the residues of the poles $z_n = i\omega_n$ to the contour integral over the small circles around these poles as shown on the left of Fig.~\ref{matsubara_sum}. 
Then the sum of the small contours are converted into the sum of the two closed contours of $l_1$ and $C_1$, and of $l_2$ and $C_2$ as shown on the right of Fig.~\ref{matsubara_sum}.
Due to the condition of Eq.~(\ref{eq_condition}), the contributions of the two semi-circles of $C_1$ and of $C_2$ vanish.

Therefore, we have the relation
\begin{align}
   &\text{Res}(F(z),z=i\omega_n)\notag\\
   =&\frac{1}{2\pi i}\left(\int_{l_1}F(z)dz+\int_{l_2}F(z)dz\right)\notag\\
   =&\frac{1}{2\pi i}\left(\int_{l_1+C_1}F(z)dz+\int_{l_2+C_2}F(z)dz\right)
   \notag\\
   =&-\sum_i\text{Res}(F(z),z=z_i).
    \label{convertion1}
\end{align}
Then finally, one arrives at the following result
\begin{align}
    \frac{1}{\beta}
    \sum_n f(z_n)=-\sum_i\text{Res}(F(z),z=z_i).
    \label{convertion2}
\end{align}
These manipulations are often called Sommerfeld-Watson transformation, the details of which are, for instance, in Refs.~\cite{Watson:1918ca,Sommerfeld:1949ca}.

\section{The calculation of one-loop two-point function at finite temperature}\label{sec:2pft}

Using Eq.~(\ref{convertion1}), the one-loop two-pont function at finite temperature can be computed as
\begin{align}
&I_2(E,m_1,m_2,T)\notag\\
=&\int\frac{d^3l}{(2\pi)^3}\int\frac{dz}{2\pi i}\frac{1}{(z^2-w_1^2+i\epsilon)[(E-z)^2-w_2^2+i\epsilon]}\notag\\
   &\times\frac{1}{2}\coth\biggl(\frac{1}{2}\beta z\biggr) \notag \\
   =&\int\frac{d^3l}{(2\pi)^3}\biggl(\text{Res}\biggl[F(z)=\frac{1}{(z^2-w_1^2+i\epsilon)[(E-z)^2-w_2^2+i\epsilon]}\notag\\
   &\times\frac{1}{2}\coth\biggl(\frac{1}{2}\beta z\biggr),z_n\biggr]\biggr) \notag \\
   =&\int\frac{d^3l}{(2\pi)^3}\biggl(\sum_i\text{Res}[F(z),z=z_i]\biggr) \nonumber \\
   =&\int\frac{d^3l}{(2\pi)^3}\Bigl(\text{Res}[F(z),w_1-i\epsilon]+\text{Res}[F(z),-w_1+i\epsilon]\notag\\
   &+\text{Res}[F(z),E-w_2+i\epsilon]+\text{Res}[F(z),E+w_2-i\epsilon]\Bigr) \nonumber \\
   =&\int\frac{d^3l}{2(2\pi)^3}\biggl[\frac{\coth(\beta (w_1-i\epsilon)/2)}{2w_1[(E-w_1)^2-w_2^2+i\epsilon]}\notag\\
   &+\frac{\coth(\beta (w_1-i\epsilon)/2)}{2w_1[(E+w_1)^2-w_2^2+i\epsilon]}+\frac{\coth(\beta (w_2-E-i\epsilon)/2)}{2w_2[(E-w_2)^2-w_1^2+i\epsilon]}\notag\\
   &+\frac{\coth(\beta (w_2+E-i\epsilon)/2)}{2w_2[(E+w_2)^2-w_1^2+i\epsilon]}\biggr].
    \label{eq:2pftfun}
\end{align}

\section{The comparison between $T^{\text{full}}$ and $T^{\text{per}}$ at finite temperature}\label{Tfull&Tper}
\begin{figure*}[t]
    \centering
    \subfigure[The temperature effect on $|T_{\pi^0\pi^0\to\pi^0\pi^0}$|.]{
    \includegraphics[width=0.45\linewidth]{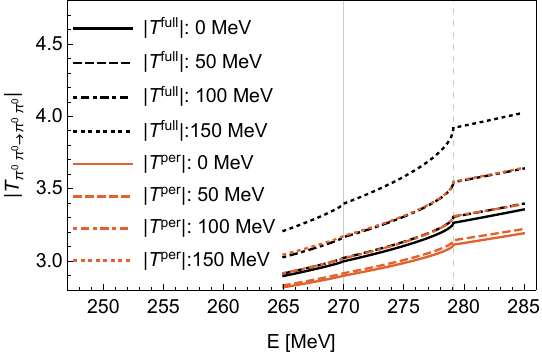}}
    \subfigure[The temperature effect on $|T_{\pi^0\pi^0\to\pi^+\pi^-}$|.]{
    \includegraphics[width=0.45\linewidth]{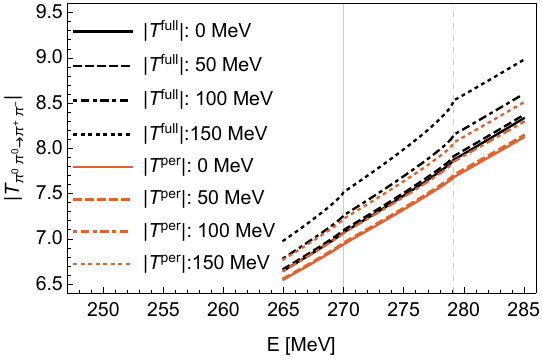}}
    \subfigure[The temperature effect on $|T_{\pi^+\pi^-\to\pi^+\pi^-}$|.]{
    \includegraphics[width=0.45\linewidth]{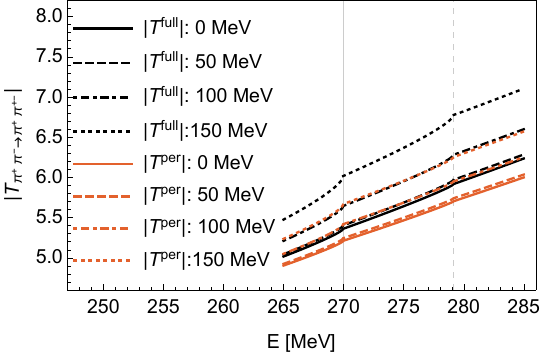}}
    \caption{   The temperature effect on each element of $|T^{\text{full}}|$ (black)  and $|T^{\text{per}}|$ (red). The solid, dashed, dot-dashed, and dotted lines are for the lineshape of the results at temperatures $T=0$ MeV, $50$ MeV, $100$ MeV, and $150$ MeV, respectively. The black solid vertical line and black dashed vertical line indicate the thresholds of $\pi^0\pi^0$ and $\pi^+\pi^-$, respectively.}
    \label{fig:Tfull&Tper}
\end{figure*}

\begin{figure*}
    \centering
    \includegraphics[width=0.45\linewidth]{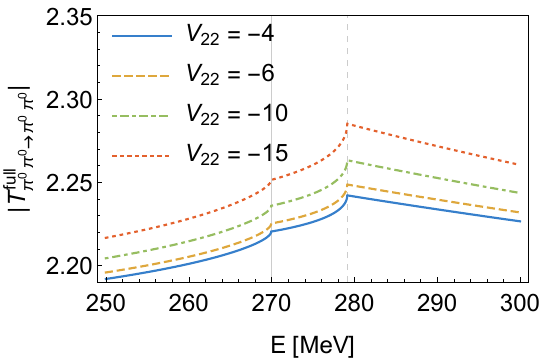}
    \caption{The $V_{22}$ effect on $|T_{\pi^0\pi^0\to\pi^0\pi^0}^{\text{full}}|$. $V_{11}=-2$, $V_{12}=V_{21}=-3$. The solid vertical gray line and dashed vertical gray line represents the $\pi^0\pi^0$ and $\pi^+\pi^+$ thresholds, respectively.}
    \label{fig:V22_effect}
\end{figure*}

Here we would like to comment on the mutual roles of finite temperature and the interaction in the transition amplitude $T$.  
Ideally, we may want to include all diagrams of $s$, $t$, $u$ and tad-poles to all orders, which however is a formidable task.  
In fact, the cusp is primarily caused  by the singularity in the s-channel process.  
Therefore, we include only the $s$-channel diagrams to all orders in the form of
$T$-matrix.  Now to see the effect of the interaction at finite temperature, we compare the $T^{\text{full}}$ and $T^{\text{pert}}$, where the former is the solution of the LSE while the latter is the one of perturbation of leading order, 
\begin{align}
T^{\text{pert}} = V + VGV.\label{eq:Tper}
\end{align}
Details with concrete examples are shown in Fig.~\ref{fig:Tfull&Tper} and Fig.~\ref{fig:V22_effect}. In Fig.~\ref{fig:Tfull&Tper} we see that the difference between $T^{\text{full}}$ and $T^{\text{pert}}$ is obvious 
at various temperatures; the amplitude $T^{\text{full}}$ is larger than $T^{\text{pert}}$ 
and so is the cusp behavior. 
Moreover, the difference between the two amplitudes increases as the temperature increases.
Fig.~\ref{fig:V22_effect} shows the interaction effect, where we see that 
the amplitude and hence cusp are enhanced as the attractive interaction 
$V_{22}$ is increased.

The interaction strengths employed there are attractive but not so strong, and there is a virtual state in the channel 2.    
In general, as the attraction is increased and the virtual state approaches the threshold, the cusp will become sharper.  
At the point where the virtual state turns into the zero-energy bound state, the hight of the cusp becomes infinite.  
As the interaction is further increased the peak moves below the threshold, and turns into a Feshbach resonance in the scattering of channel 1.  
The singularity still remains at the threshold where the derivative becomes minus infinity.  
For the pion scattering, to realize such a sharp peak, an unrealistically large attraction 
is needed.  In reality the cusp of the pion is rather small, though it exists.

 
\section{The calculation of the isospin breaking part in the propagators of pion and charmed meson system}\label{sec:isobreaking}
The $\pi\pi$ propagator including isospin symmetry breaking from the position 0 to $x$ is calculated as
\begin{widetext}
\begin{align}
    &~G(x,0;T)^{\pi\pi}\notag\\
    =&~\langle 0|[\pi(x)\pi(x)]_{I=0}[\pi(0)\pi(0)]_{I=2}|0\rangle\notag\\
    =&~\langle 0|(\sqrt{\frac{1}{3}}\pi^+(x)\pi^-(x)-\sqrt{\frac{1}{3}}\pi^0(x)\pi^0(x)+\sqrt{\frac{1}{3}}\pi^-(x)\pi^+(x))(\sqrt{\frac{1}{6}}\pi^+(0)\pi^-(0)+\sqrt{\frac{2}{3}}\pi^0(0)\pi^0(0)+\sqrt{\frac{1}{6}}\pi^-(0)\pi^+(0))|0\rangle\notag\\
    =&~\frac{\sqrt{2}}{3}\langle 0|\pi^+(x)\pi^-(x)\pi^+(0)\pi^-(0)|0\rangle-\frac{\sqrt{2}}{3}\langle 0|\pi^0(x)\pi^0(x)\pi^0(0)\pi^0(0)|0\rangle\notag\\
    =&~ \frac{\sqrt{2}}{3}[I_2(E,m_{\pi^+},m_{\pi^-},T)-I_2(E,m_{\pi^0},m_{\pi^0},T)].
\end{align}
\end{widetext}
The $D\bar{D}^\ast$ propagator including isospin symmetry breaking from the position 0 to $x$ is calculated as
\begin{widetext}
\begin{align}
    &~G(x,0;T)^{D\bar{D}^*}\notag\\
    =&~\langle 0|[D(x)\bar{D}^*(x)]_{I=0}[D(0)\bar{D}^*(0)]_{I=1}|0\rangle\notag\\
    =&~\langle 0|\sqrt{\frac{1}{2}}(D^0(x)\bar{D}^{*0}(x)-D^+(x)D^{*-}(x))\sqrt{\frac{1}{2}}(D^0(0)\bar{D}^{*0}(0)+D^+(0)D^{*-}(0))|0\rangle\notag\\
    =&~\frac{1}{2}(\langle 0|D^0(x)\bar{D}^{*0}(x)D^0(0)\bar{D}^{*0}(0)|0\rangle-\langle 0|D^+(x)D^{*-}(x)D^+(0)D^{*-}(0)|0\rangle)\notag\\
     =&~\frac{1}{2}\biggl[I_2\biggl(E,m_{D^0}(T)-\frac{i\Gamma_{D^0}(T)}{2},m_{\bar{D}^{*0}}(T)-\frac{i\Gamma_{\bar{D}^{*0}}(T)}{2},T\biggr)-I_2\biggl(E,m_{D^+}(T)-\frac{i\Gamma_{D^+}(T)}{2},m_{D^{*-}}(T)-\frac{i\Gamma_{D^{*-}}(T)}{2},T\biggr)\biggr],
\end{align}
\end{widetext}
where the temperature dependence of cahrmed meson mass and width, i.e., $m(T)$ and $\Gamma(T)$, is considered, because the charmed mesons are unstable against the strong decay at finite temperature. 

\end{document}